\documentclass[journal]{IEEEtran}

\ifCLASSINFOpdf
 
\else
 
\fi

\usepackage{orcidlink}
\usepackage{multirow}
\usepackage{amsmath}
\usepackage{amsfonts}
\usepackage{amssymb}
\usepackage{xcolor}
\usepackage{flushend}
\usepackage{graphicx}
\usepackage{algpseudocode}

\usepackage{textcomp}

\usepackage{multicol}
\usepackage[caption=false,font=footnotesize]{subfig}
\usepackage{algorithm}

\usepackage{cite}

\begin{document}

\title{System-Aware Adaptive CSI Feedback via RL-Guided Autoencoder Switching in Multi-User MIMO Systems}

\author{Maryam Ansarifard \orcidlink{0000-0002-6305-1484},
        Mohit K. Sharma,
        George Exarchakos \orcidlink{0000-0002-5773-0720},
        Kishor C. Joshi \orcidlink{0000-0002-6047-5812}
\thanks{This work was supported by the Dutch National Growth Fund
project ``6G Future Network Services (FNS)'' and the Eindhoven
Hendrik Casimir Institute (EHCI), The Netherlands.}%
\thanks{M. Ansarifard, G. Exarchakos, and K. C. Joshi are with the Department of Electrical Engineering, Eindhoven University of Technology, Eindhoven, The Netherlands. E-mails: (\{M.Ansarifard, G.Exarchakos, K.C.Joshi\}@tue.nl).}%
\thanks{M. K. Sharma is with the Directed Energy Research Center, Technology Innovation Institute, Abu Dhabi, UAE. E-mail: Mohit.Sharma@tii.ae.}}


\maketitle
\begin{abstract}
This paper proposes a system-aware adaptive channel state information (CSI) feedback framework for massive multiple-input multiple-output (mMIMO) systems, aiming to dynamically optimize the trade-off between reconstruction fidelity and signaling overhead. While deep learning-based autoencoders (AEs) have enabled significant CSI compression, conventional fixed-ratio schemes fail to adapt effectively to non-stationary channel conditions. To address this limitation, we develop a reinforcement learning (RL)-driven control framework that operates over a bank of pretrained multi-rate AEs, each corresponding to a distinct compression ratio (CR). At each time step, a centralized RL agent selects the most suitable CR for each user based on observed channel conditions and system performance indicators.
Distinct from conventional mean squared error (MSE)-centric designs, we introduce a system-aware reward formulation that jointly accounts for spectral efficiency through the signal-to-interference-plus-noise ratio (SINR), CSI feedback overhead, and AE-switching cost. Simulation results on high-dimensional delay-domain CSI datasets demonstrate that the proposed RL-guided framework effectively balances the overhead--accuracy tradeoff and adapts to dynamic channel environments. The proposed method improves spectral efficiency and feedback efficiency compared with fixed compression schemes and adaptive baselines, while maintaining a modest computational and memory footprint. Averaged over different numbers of users and across all considered baselines, the proposed RL framework reduces the CSI feedback cost by more than $53.4\%$, improves the average downlink sum rate by $53.64\%$, and reduces the normalized mean squared error (NMSE) by $22.38\%$. These results demonstrate its ability to achieve a more efficient rate--accuracy--feedback tradeoff under dynamic wireless conditions.
\end{abstract}


\IEEEpeerreviewmaketitle

\section{Introduction}
\IEEEPARstart{M}{assive} multiple-input multiple-output (mMIMO) has emerged as a fundamental technology for 5G-Advanced and 6G networks, offering substantial gains in spectral efficiency and spatial multiplexing. To fully realize these benefits, particularly beamforming and multi-user interference nulling, the base station (BS) requires accurate downlink channel state information (CSI). In practical frequency-division duplexing (FDD) systems, the acquisition of downlink CSI starts at the user equipment (UE) through pilot-based channel estimation \cite{11131223}. Since uplink and downlink operate over different frequency bands in FDD, channel reciprocity cannot be directly exploited at the BS. Therefore, the BS periodically (or on demand) transmits reference signals, called channel state information reference signals (CSI-RS), across time and frequency resources. Using these known pilot symbols, each UE estimates the downlink channel over the occupied OFDM subcarriers. These estimates capture the spatial-frequency characteristics of the propagation environment, including path loss, multipath fading, and scattering effects. Once the channel is estimated, the UE must convey this information back to the BS to enable precoding and beamforming. However, directly feeding back the full CSI matrix is infeasible in large-scale MIMO systems due to the enormous dimensionality of the channel representation. This challenge becomes more pronounced as the number of antennas, subcarriers, and users increases, leading to significant feedback overhead that limits system scalability.

Traditional CSI feedback methods, including codebook-based quantization \cite{ziao2025review, suarez20233gpp}, compressive sensing \cite{article}, and low-rank modeling \cite{9376994, he2024low}, provide partial relief by exploiting structural properties of the channel such as sparsity or spatial correlation. However, these approaches often rely on handcrafted models and linear assumptions that struggle to capture the highly complex and non-linear characteristics of wireless propagation environments. Furthermore, their performance typically degrades under dynamic channel conditions, and some methods incur substantial computational overhead or reconstruction latency, limiting their practical deployment in large-scale systems. In addition, these techniques often require careful parameter tuning and prior knowledge of channel statistics, which may not be readily available or may vary significantly across different scenarios and frequency bands. As a result, their robustness and generalization capability remain limited, particularly in heterogeneous networks with diverse user mobility patterns and propagation conditions. Moreover, the trade-off between feedback compression and reconstruction accuracy is difficult to balance, as aggressive compression can lead to significant information loss, ultimately degrading beamforming and system performance. These limitations highlight the need for more flexible and data-driven approaches that can efficiently learn compact representations of CSI while adapting to complex and time-varying wireless environments.

Recent research has increasingly explored data-driven approaches, particularly machine learning (ML) techniques, for CSI compression and reconstruction. Unlike model-driven methods, ML-based approaches can learn implicit channel representations directly from data without relying on rigid analytical assumptions. Among these techniques, deep learning (DL)-based autoencoders (AEs) have emerged as a promising solution for CSI feedback in mMIMO systems \cite{9417115, 10465185, guo2022overview, cui2024lightweight}. These architectures consist of an encoder network deployed at the UE and a decoder network located at the BS. The encoder compresses the high-dimensional CSI matrix into a compact latent representation, which is then transmitted through the feedback link, while the decoder reconstructs the CSI at the BS. By jointly optimizing both networks in an end-to-end manner, AE-based frameworks can effectively capture complex spatial–frequency channel correlations and achieve significantly higher compression efficiency compared to conventional approaches.

Despite their success, nearly all existing AE-based schemes operate with a fixed compression ratio (CR) determined offline. This static-ratio approach is fundamentally flawed given the non-stationary nature of 6G environments. A fixed CR cannot adapt to varying channel realizations, fluctuations in signal-to-noise ratio (SNR), or changing application-specific accuracy requirements, often resulting in either wasted feedback resources or insufficient CSI fidelity. In practical 6G scenarios, user mobility, scattering evolution, and blockages cause rapid CSI dynamics. While a user might initially be assigned a specific CR based on long-term features, channel distribution drift can quickly render that ratio suboptimal. In practice, a user’s CSI requirements evolved based on local scattering, mobility-induced Doppler shifts, and sudden blockages (LoS to NLoS transitions). Utilizing a high compression level (e.g., $1/32$) during low-SNR conditions may result in beamforming misalignment, while a low compression level (e.g., $1/4$) during stable periods wastes valuable feedback bits. Furthermore, a single pretrained AE may become mismatched to changing channel conditions. Although online fine-tuning can alleviate this mismatch, it requires backpropagation and gradient storage at the UE and may also create inconsistencies between the encoder and decoder. To avoid these costs, our framework keeps all AE parameters frozen during deployment and adapts the feedback configuration by switching among pretrained CR-specific AEs.

To address these limitations, we move beyond conventional CSI reconstruction objectives and propose a system-aware adaptive feedback framework that explicitly considers network-level performance and resource constraints. Instead of optimizing CSI compression solely for reconstruction accuracy, our approach dynamically adapts the feedback process based on the current channel conditions and system requirements. We formulate the CSI feedback control problem as a Markov decision process (MDP), where a centralized reinforcement learning (RL) agent orchestrates the feedback strategy across users. At each decision step, the agent observes the system state, which captures key indicators of channel quality, feedback cost, and reconstruction performance, including the normalized mean squared error (NMSE), defined as the normalized distortion between the original CSI and the reconstructed CSI at the BS, and selects a pretrained AE from a multi-rate AE bank, which effectively results in choosing a specific CR. The selected AE is used for encoding the CSI into latent space, which is then used at the BS for reconstruction using the corresponding decoder. 

Through this formulation, the RL agent learns to balance three conflicting objectives: maximizing spectral efficiency through accurate CSI reconstruction, minimizing feedback overhead on the uplink control channel, and limiting unnecessary switching among compression-specific pretrained AEs. The proposed framework therefore allocates feedback resources according to the evolving channel conditions while keeping all AE parameters frozen during online operation.

Compared with static-ratio AE-based schemes, the proposed approach provides adaptive and context-aware CSI feedback, allowing the system to react to channel distribution shifts, variations in SNR, and evolving propagation environments. As a result, the framework improves overall system efficiency while avoiding unnecessary feedback transmissions, online AE retraining, and model-parameter exchange, making it particularly suitable for highly dynamic 6G scenarios.

\subsection{Related Works}
Table~\ref{tab:literature_comparison} provides an overview of representative CSI feedback and compression methods reported in the literature and compares them with the proposed framework.
Early learning-based CSI compression methods employ convolutional or fully connected AEs to exploit spatial–frequency correlations in OFDM MIMO channels. CsiNet \cite{csinet} and its variants \cite{csinet+,csinet+dnn} demonstrate strong reconstruction performance at fixed CRs, while later works extend to attention modules, temporal correlation learning, and transformer-based latent representations \cite{csiformer, ansarifard2025csi, caiattention, transnet, lou2025attention, 11482610}. Although more aggressive compression reduces feedback overhead,
it also increases reconstruction distortion, motivating ratio-adaptive approaches.
Prior studies comparing model-based codebook approaches with data-driven CSI compression methods, including convolutional and transformer-based models, report average throughput gains of approximately 5\%--15\% in low-feedback-overhead scenarios. As the feedback overhead increases, the gain decreases to around 3\%--9\% in medium-overhead settings~\cite{chen2025csi}.

Several studies introduce heuristic or supervised methods to assign different CSI CRs across users or time. For instance, the deep AE-based CSI compression framework in \cite{10706622} shows that runtime complexity is largely independent of the CR, making adaptive compression feasible. However, CRs are selected offline and remain static during deployment, which limits robustness under channel distribution drift caused by user mobility or environmental changes. Similarly, \cite{10012898} proposes an adaptive DNN-based CSI feedback scheme that selects the CR via a classification block and improves NMSE under more realistic clustered delay line (CDL) channel models. While this method adapts CRs, the adaptation is rule-based, without explicitly considering system-level performance metrics such as throughput loss or feedback overhead trade-offs over time.

User-driven adaptive feedback approaches, such as the nested-dropout and vector-quantization-based scheme in \cite{10208156}, enable variable-length CSI feedback using a shared codebook. These methods efficiently accommodate heterogeneous user requirements and outperform fixed-rate baselines under varying feedback budgets. However, they do not address distribution drift nor update the encoder/decoder representations over time. Likewise, PFnet with adaptive compression in \cite{10967811} dynamically adjusts CRs based on SNR and focuses on pilot compression rather than CSI compression. While this reduces overhead and improves NMSE, adaptation decisions rely on instantaneous channel quality indicators and remain myopic, without learning temporal adaptation policies or accounting for computational retraining costs. 

Exploiting side information to guide adaptive feedback is also being considered in literature. For example, \cite{8090842} leverages user location correlation to reduce feedback overhead and improve MU-MIMO throughput. Although Location-based CSI feedback compression (LFC) scales well with the number of users, it relies on explicit location information and predefined grouping strategies, limiting applicability in dynamic or privacy-constrained environments.

Cross-layer adaptive feedback compression in \cite{10.1145/2500423.2500446} dynamically adjusts CSI compression intensity using practical metrics extracted from packet preambles and demonstrates strong experimental performance. However, this relies on handcrafted adaptation metrics and does not leverage learned CSI representations or online model refinement.

\subsection{Our contribution}

We consider a closed-loop CSI feedback framework in which the BS hosts the RL agent and a bank of pretrained autoencoder decoders, while the corresponding encoders are available at the UEs. At each transmission interval, the BS selects the operating CR based on the observed network state and communicates the corresponding encoder index to the UE through lightweight control signaling. The UE compresses its CSI using the selected encoder and feeds the compressed representation back to the BS for reconstruction and downlink transmission. Within this setting, the main contributions of this work are summarized as follows:

\begin{itemize}

\item We propose a drift-resilient CSI compression framework for multi-user MIMO systems based on RL-guided switching among pretrained autoencoders with different CRs. The proposed framework adapts CSI feedback to changing channel conditions without requiring online retraining or fine-tuning during deployment.

\item We formulate adaptive CSI compression as a sequential decision-making problem and develop a Dueling-DQN-based controller that dynamically selects the operating CR according to the observed network state, including reconstruction quality, signal-to-interference-plus-noise ratio (SINR), and feedback conditions.

\item We design a system-aware reward function that jointly accounts for CSI reconstruction distortion, feedback overhead, switching cost, and achievable downlink sum rate. Consequently, the proposed framework optimizes CSI compression from a system-level perspective rather than focusing solely on reconstruction accuracy.

\item We conduct extensive simulations under diverse channel scenarios and user densities. The results demonstrate that the proposed framework consistently outperforms rule-based, supervised-learning, and fixed-CR baselines in terms of reconstruction quality, adaptation robustness, communication performance, and long-term reward.

\end{itemize}

To the best of our knowledge, no existing work explicitly enables an RL agent to dynamically select among pretrained autoencoders with different CRs in a multi-user MIMO CSI compression framework. The proposed framework enables autonomous and context-aware CSI feedback adaptation under dynamic channel conditions. This provides a scalable step toward “Agentic AI” in the physical layer, where communication systems can intelligently optimize feedback strategies in response to evolving wireless environments and network requirements.
\begin{table*}[t]
\centering
\caption{Comparison of representative CSI feedback and compression approaches.}
\label{tab:literature_comparison}
\renewcommand{\arraystretch}{1.2}
\resizebox{\textwidth}{!}{
\begin{tabular}{|l|c|c|c|c|c|}
\hline
\textbf{Method}
&
\multicolumn{3}{c|}{\textbf{Adaptation Capabilities}}
&
\multicolumn{1}{c|}{\textbf{Optimization Objectives}}
&
\textbf{Deployment}
\\
\cline{2-6}

&
\textbf{Adaptive} &
\textbf{Online} &
\textbf{Drift} &
\textbf{System} &
\textbf{MU}
\\

&
\textbf{Compression} &
\textbf{Adaptation} &
\textbf{Aware} &
\textbf{Aware} &
\textbf{Support}
\\
\hline

CsiNet/ CsiNet+ / CsiNet-DNN~\cite{csinet,csinet+,csinet+dnn}
& $\times$ & $\times$ & $\times$ & $\times$ & $\triangle$ \\
\hline

Transformer-based CSI feedback~\cite{csiformer,caiattention,transnet,lou2025attention,11482610}
& $\times$ & $\times$ & $\times$ & $\times$ & $\triangle$ \\
\hline

Adaptive DNN CSI feedback~\cite{10012898}
& $\checkmark$ & $\triangle$ & $\times$ & $\times$ & $\triangle$ \\
\hline

Runtime-adaptive AE compression~\cite{10706622}
& $\checkmark$ & $\triangle$ & $\times$ & $\times$ & $\triangle$ \\
\hline

Nested Dropout / VQ methods~\cite{10208156}
& $\checkmark$ & $\checkmark$ & $\times$ & $\times$ & $\checkmark$ \\
\hline

PFNet~\cite{10967811}
& $\checkmark$ & $\checkmark$ & $\times$ & $\triangle$ & $\triangle$ \\
\hline

Location-based and cross-layer methods~\cite{8090842,10.1145/2500423.2500446}
& $\checkmark$ & $\checkmark$ & $\triangle$ & $\checkmark$ & $\checkmark$ \\
\hline

\textbf{Proposed Method}
& $\checkmark$ & $\checkmark$ & $\checkmark$ & $\checkmark$ & $\checkmark$ \\
\hline

\end{tabular}}
\vspace{1mm}

\footnotesize
$\checkmark$: supported; 
$\triangle$: partially supported; 
$\times$: not supported.
\end{table*}

\subsection{Paper Organization}
The rest of this paper is organized as follows. The system model is presented in Section~\ref{System Model}. In Section~\ref{Solution}, the problem formulation and the solution to the problem are provided. The evaluation setup and performance analysis are presented in Sections~\ref{result} and~\ref{analysis}, respectively, and Section~\ref{conclusion} concludes the paper.
\subsubsection*{Notations}
Lowercase bold symbols denote vectors and uppercase bold symbols denote matrices. The operators \((\cdot)^T\), \((\cdot)^H\), \(\|\cdot\|_2\), and \(\mathbb{E}\{\cdot\}\) denote transpose, Hermitian transpose, Euclidean norm, and expectation, respectively.

\section{System Model} \label{System Model}
\begin{figure*}[t]
	\centering
	\includegraphics[width=\textwidth]{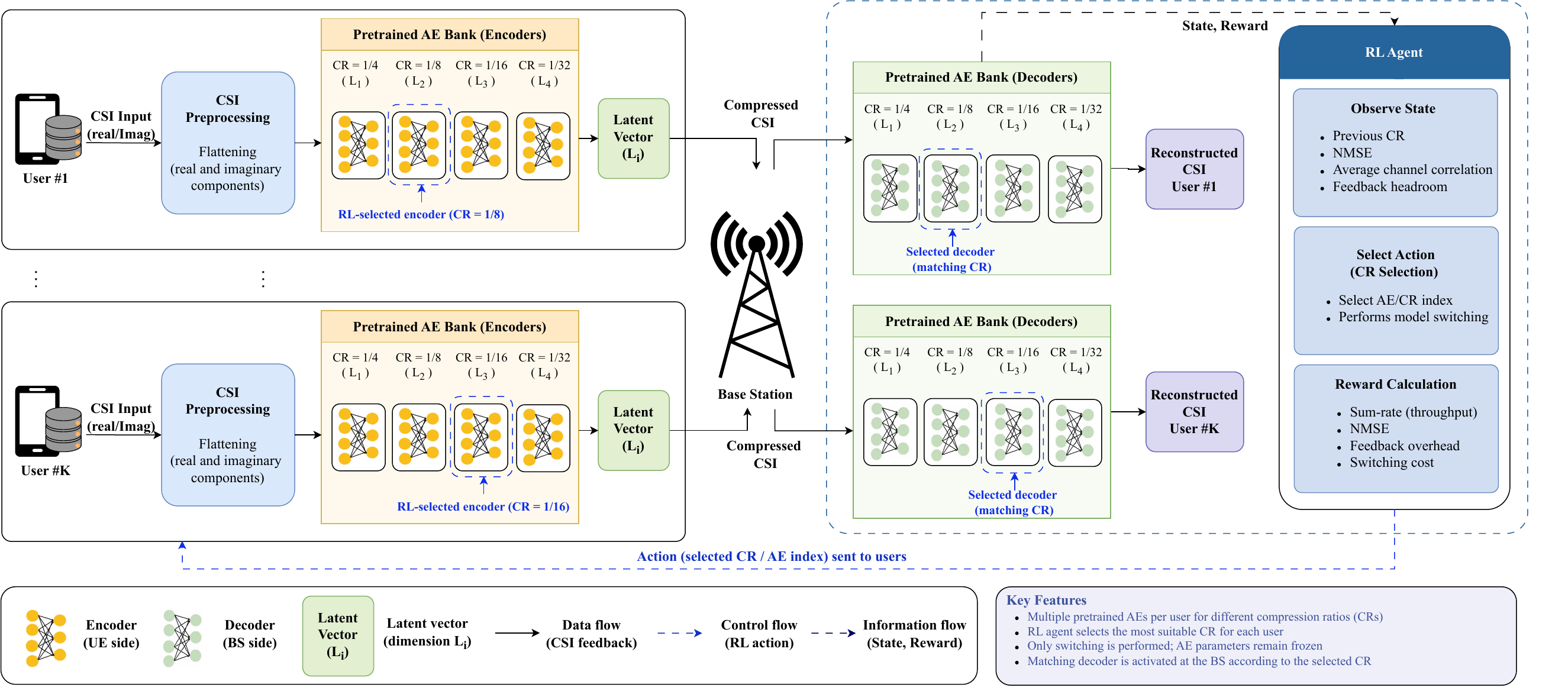}
	\caption{System architecture and workflow of the proposed adaptive CSI compression framework.}
\end{figure*}
We consider a downlink multi-user MIMO (MU-MIMO) orthogonal frequency division multiplexing (OFDM) system where the BS is equipped with $N_t$ transmit antennas\footnote{The term antenna in this paper refers to an antenna port, which can be mapped to multiple physical antennas.} and serves $K$ mobile single-antenna users over $\tilde{N}_s$ OFDM subcarriers across $T$ time slots.
In practical systems, the UE estimates the downlink channel using pilot signals transmitted from the BS \cite{gebre2021comparative}. Let $\mathbf{x}_n[t] \in \mathbb{C}^{N_t}$ denote the pilot symbol vector transmitted by the BS on the $n$th subcarrier at time slot $t$. The received signal at user $k$ can then be expressed as
\begin{equation}
y_{k,n}[t] = \mathbf{h}_{k,n}^H[t]\mathbf{x}_n[t] + z_{k,n}[t],
\end{equation}
where $\mathbf{h}_{k,n}[t] \in \mathbb{C}^{N_t}$ denotes the downlink channel vector between the BS and user $k$ on the $n$th subcarrier at time slot $t$, and $z_{k,n}[t] \sim \mathcal{CN}(0,\sigma^2)$ represents the additive white Gaussian noise (AWGN).

The downlink spatial-frequency channel matrix of user $k$ at time slot $t$ is denoted as
\begin{equation}
\mathbf{\tilde{H}}_k[t] =
\begin{bmatrix}
\mathbf{h}_{k,1}^H[t] \\
\mathbf{h}_{k,2}^H[t] \\
\vdots \\
\mathbf{h}_{k,\tilde{N}_s}^H[t]
\end{bmatrix}
\in \mathbb{C}^{\tilde{N}_s \times N_t}.
\end{equation}

By applying a two-dimensional discrete Fourier transform (2D-DFT), the channel can be represented in the angular-delay domain as
\begin{equation}
\mathbf{H}_k = \mathbf{F}_d \, \mathbf{\tilde{H}}_k \, \mathbf{F}_a^{\mathrm{H}},
\end{equation}
where $\mathbf{F}_d \in \mathbb{C}^{\tilde{N_s} \times \tilde{N_s}}$ and $\mathbf{F}_a \in \mathbb{C}^{N_t \times N_t}$ denote the normalized DFT matrices in the delay and angular domains, respectively. Their $(m,n)$-th entries are given by
\begin{align}
[\mathbf{F}_d]_{m,n}
=
\frac{1}{\sqrt{\tilde N_s}}
e^{-j\frac{2\pi mn}{\tilde N_s}},
\quad
m,n=0,\ldots,\tilde N_s-1,\\\nonumber
[\mathbf{F}_a]_{m,n}
=
\frac{1}{\sqrt{N_t}}
e^{-j\frac{2\pi mn}{N_t}},
\quad
m,n=0,\ldots,N_t-1.
\end{align}

The transformation concentrates most of the channel energy into a limited number of angular-delay components, thereby providing a sparse representation that is well suited for CSI compression.

Due to the limited delay spread of multipath propagation,
the energy of $\mathbf{H}_k$ in the delay domain is predominantly concentrated
in its first $N_\tau \leq \widetilde{N}_s$ rows, while the remaining rows
contain negligible values. Exploiting this sparsity, only the first $N_\tau$
delay-domain rows are retained to form a truncated channel matrix. With a
slight abuse of notation, we denote the resulting truncated matrix by
$\mathbf{H}_k \in \mathbb{C}^{N_\tau \times N_t}$. The symbol $N_s$ is used
later to denote the number of retained/evaluated subcarriers over which
per-subcarrier quantities such as $\text{SINR}_{k,n}$ and $\mathbf{W}_n$
are defined. Furthermore, to facilitate real-valued processing, the real and imaginary parts of $\mathbf{H}_k$ are concatenated, yielding a real-valued representation $\mathbf{H}_k \in \mathbb{R}^{2N_s \times N_t}$.


\subsection{Autoencoder-Based CSI Compression}
Each user compresses CSI using the encoder model in pre-trained autoencoder and transmits a compressed CSI latent vector:
\begin{equation}
\mathbf{z}_k[t] = f_{\text{enc}}(\mathbf{H}_k[t]; r^{\text{CR}}_k[t]),
\end{equation}
where $f_{\text{enc}}(\cdot)$ is the encoder network, and $r^{\text{CR}}_k[t] \in \mathcal{R} = \{1/4, 1/8, 1/16, 1/32\}$ is the CR selected for user $k$ in slot $t$. At the BS, decoder model of pre-trained autoencoder is used to reconstruct CSI for each user as:
\begin{equation}
\widehat{\mathbf{H}}_k[t] = f_{\text{dec}}(\mathbf{z}_k[t]).
\end{equation}

The CSI reconstruction quality for each user is measured using the NMSE:

\begin{equation}
\text{NMSE}_k[t] = \frac{ \| \mathbf{H}_k[t] - \widehat{\mathbf{H}}_k[t] \|_2^2 }{ \| \mathbf{H}_k[t] \|_2^2 }.
\end{equation}

Let $D=2N_sN_t$ denote the dimension of the real-valued channel representation at the encoder input. The latent-vector length corresponding to the selected compression ratio is
$L_k[t]=D r_k^{\mathrm{CR}}[t]=2N_sN_t r_k^{\mathrm{CR}}[t]$.
These latent values must be communicated from the UE to the BS through the uplink CSI feedback channel. However, the elements of $\mathbf{z}_k[t]$ are real-valued numbers, not bits. Before digital transmission, each latent element must be quantized. Assuming $q$ bits per latent element, the CSI feedback payload is
\begin{equation}\label{bits}
B_k[t] =qL_k[t].
\end{equation}

\subsection{Precoding and SINR Calculation}
\subsubsection{Precoding}
To evaluate the MU-MIMO system performance, we compute the SINR by applying the precoders designed from the reconstructed CSI to the ground-truth channel $\mathbf{h}_{k,n}$. The total transmit power $P_{\mathrm{tx}}$ is divided equally among the $K$ users, such that the power per user is $p = P_{\mathrm{tx}}/K$. The receiver noise is modeled as AWGN with power $\sigma^2$.

The BS employs a linear precoding matrix $\mathbf{W}_n = [\mathbf{w}_{1,n}, \dots, \mathbf{w}_{K,n}] \in \mathbb{C}^{N_t \times K}$ for each subcarrier $n$. The unnormalized precoding vectors $\mathbf{v}_{k,n}$ are determined based on the chosen strategy. We employ zero-forcing (ZF)  to mitigate inter-user interference by inverting the estimated channel matrix $\widehat{\mathbf{H}}_n = [\widehat{\mathbf{h}}_{1,n}^T, \dots, \widehat{\mathbf{h}}_{K,n}^T]^T \in \mathbb{C}^{K \times N_{t}}$. The unnormalized matrix $\mathbf{V}_n$ is given by:
    \begin{equation}
    \mathbf{V}_n = \widehat{\mathbf{H}}_n^{H} \left( \widehat{\mathbf{H}}_n \widehat{\mathbf{H}}_n^{H} \right)^{-1}.
    \end{equation}

Each precoding vector is normalized to unit norm to maintain strict power allocation:
\begin{equation}
\mathbf{w}_{k,n} = \frac{\mathbf{v}_{k,n}}{\|\mathbf{v}_{k,n}\|}.
\end{equation}
\subsubsection{SINR Calculation}
The instantaneous SINR for user $k$ on subcarrier $n$ captures the degradation caused by the mismatch between the estimated precoder $\mathbf{w}_{k,n}$ and the true channel $\mathbf{h}_{k,n}$:
\begin{equation}
\mathrm{SINR}_{k,n}
=
\frac{
p
\left|
\mathbf{h}_{k,n} \mathbf{w}_{k,n}
\right|^2
}{
\sum\limits_{k^{\prime} \neq k}
p
\left|
\mathbf{h}_{k,n} \mathbf{w}_{k^{\prime},n}
\right|^2
+
\sigma^2
}.
\end{equation}

The average SINR for user $k$ over the $N_s$ retained subcarriers is defined as:
\begin{equation}
\mathrm{SINR}_{k} = \frac{1}{N_{s}} \sum_{n=1}^{N_{s}} \mathrm{SINR}_{k,n}.
\end{equation}

Based on the variables defined in this section, the adaptive CSI feedback problem can be interpreted as selecting, at each time slot $t$, the CR $r^{\mathrm{CR}}_k[t]$ and the corresponding pretrained AE for each user $k$. This decision affects the reconstructed CSI, the reconstruction distortion, the feedback overhead, and ultimately the achieved $\mathrm{SINR}_k[t]$ after precoding. Since these quantities evolve over time due to channel variations, user mobility, and multi-user interference, the problem naturally forms a sequential decision-making process. Therefore, it can be modeled as an MDP, and an RL-based approach is suitable for learning a policy that dynamically selects the appropriate pretrained AE under time-varying communication and feedback conditions.


\section{Reinforcement Learning Formulation for Pretrained AE Switching} \label{Solution}
RL agent is employed to jointly determine the appropriate CR and the corresponding pretrained AE selection for each user. The environment is inherently stochastic due to channel variations, user mobility, multi-user interference, and time-varying propagation conditions. These dynamics are naturally captured within the MDP framework, where the agent observes a compact state representation that reflects both communication-level metrics (e.g., SINR and channel correlation) and CSI reconstruction indicators (e.g., NMSE and feedback budget conditions). Based on the observed state, the agent selects an action corresponding to one of the available pretrained multi-rate AEs, thereby enabling adaptive CSI compression through intelligent AE switching.

A key advantage of RL is its ability to learn long-term tradeoffs among competing objectives. Selecting a larger latent representation can improve CSI reconstruction but increases feedback overhead, whereas more aggressive compression reduces the feedback payload at the expense of CSI accuracy. Frequent changes in the selected CR may also introduce control and switching overhead. By optimizing a cumulative reward that combines throughput, reconstruction distortion, feedback cost, and switching cost, the agent learns a communication-aware AE-selection policy.

Moreover, the multi-user setting introduces coupling through shared feedback constraints. Rather than solving a combinatorial optimization problem, this coupling is incorporated into the reward design, enabling coordinated and scalable decision-making across users. The proposed formulation is detailed next through the definition of the state space, action space, and reward function.
\subsubsection{State Space} 
At each time slot $t$ and for each user $k$, the agent observes the state vector:
\begin{align}\label{state}
s_k[t] = &\big[ r^{\text{CR}}_k[t-1],\, \text{NMSE}_k[t-1], \text{SINR}_k[t],\, \bar{\rho}_k[t-1],b_k[t] \big], 
\end{align}
where:
\begin{itemize}
    \item $r^{\text{CR}}_k[t-1]$ denotes the CR selected for user $k$ in the previous time slot, which also determines the active pretrained AE used for CSI compression and reconstruction,
    \item $\text{NMSE}_k[t-1]$ is the reconstruction distortion from the previous time slot,
    \item $\bar{\rho}_k[t-1]$ denotes the average channel correlation between user $k$ and the remaining co-scheduled users from the previous time slot, defined as
    \begin{equation}
    \bar{\rho}_k[t-1]
    =
    \frac{1}{K-1}
        \sum_{\substack{j=1 \\ j \neq k}}^{K}
        \frac{\left| \mathbf{h}_k^{\mathrm{H}}[t-1]\mathbf{h}_j[t-1] \right|}
        {\left\|\mathbf{h}_k[t-1]\right\|_2 \left\|\mathbf{h}_j[t-1]\right\|_2},
    \end{equation}
    where $\mathbf{h}_k[t-1]$ denotes the estimated channel vector of user $k$ of previous time slot.
   \item \(b_k[t]\) represents the normalized remaining feedback-budget
    headroom available to user \(k\), computed using the normalized
    feedback costs in \eqref{eq:normalized_feedback}.
\end{itemize}

This state representation is deliberately low-dimensional and normalized, while preserving the main dependencies among channel dynamics, multi-user interference, CSI reconstruction quality, and feedback-resource allocation. Since the pretrained AE parameters remain frozen during online operation, AE weights and optimizer states need not be included in the state. The environment remains stochastic because of time-varying channels, user mobility, multi-user interference, and transitions among propagation conditions, which are naturally represented within the MDP framework.
\subsubsection{Action Space}
At each decision step, the RL agent selects an action:
\begin{equation}
a_k[t] \in \{0,1,\ldots,|\mathcal{R}|-1\},
\end{equation}
where each action corresponds to selecting one of the available pretrained autoencoders associated with a specific CR in the predefined set $\mathcal{R}$.
More specifically, action $a_k[t]=i$ activates the pretrained AE corresponding to CR $\mathcal{R}_i$ for user $k$ at time slot $t$. 

\subsubsection{Reward Function Design}

In line with 3GPP practice \cite{3gpp_tr38843_ai_nr_2024}, the impact of CSI reconstruction quality is evaluated through its effect on downlink communication performance. Nevertheless, CSI reconstruction distortion and post-precoding SINR provide complementary information. NMSE measures the fidelity of the reconstructed CSI, whereas SINR captures the downstream effect of the reconstructed CSI on multi-user precoding, interference suppression, and spectral efficiency. Since the mapping from CSI distortion to SINR is nonlinear and depends on channel geometry, user correlation, and the spatial structure of the reconstruction error, the two metrics are not redundant. The reward function is therefore designed to be system-aware, jointly balancing downlink throughput, CSI reconstruction fidelity, feedback overhead, and CR switching cost. At each time slot $t$, the per-user reward for user $k$ is defined as
\begin{align} \label{reward}
    {r}_k[t] &= \delta \log_2\!\bigl(1 + \mathrm{SINR}_k[t]\bigr) -
\alpha \log\!\bigl(1 + \mathrm{NMSE}_k[t]\bigr)\\\nonumber
& -
\beta \widetilde{B}_k[t] -
c_{\mathrm{sw}}\mathbb{I}\{r_k^{\mathrm{CR}}[t]\neq r_k^{\mathrm{CR}}[t-1]\},
\end{align}
where $\delta$, $\alpha$, $\beta$, and $c_{\mathrm{sw}}$ are positive weighting coefficients.

The first term represents the achievable downlink rate obtained from the post-processing SINR, thereby encouraging actions that improve system throughput. The second term penalizes poor CSI reconstruction quality through the NMSE, using a logarithmic scaling to prevent excessive sensitivity to outliers. The third term accounts for feedback overhead. We normalize the feedback payload with respect to the largest candidate latent
representation. Let
\begin{equation}
B_{\mathrm{ref}}
=
qD r_{\max}^{\mathrm{CR}},
\qquad
r_{\max}^{\mathrm{CR}}=\frac{1}{4}.
\end{equation}
The normalized feedback cost of user \(k\) is then
\begin{equation}
\label{eq:normalized_feedback}
\widetilde{B}_k[t]
=
\frac{B_k[t]}{B_{\mathrm{ref}}}
=
\frac{r_k^{\mathrm{CR}}[t]}{r_{\max}^{\mathrm{CR}}}.
\end{equation}
The final term in~\eqref{reward} imposes a fixed cost when the agent triggers a CR change and switching to the autoencoder pretrained on the chosen CR.

To enforce a system-level feedback constraint in the multi-user setting, an additional penalty is applied when the total feedback overhead exceeds a predefined budget. Let 
\begin{equation}\label{total cost}
    \widetilde{B}[t]
=
\sum_{k=1}^{K}
\widetilde{B}_k[t],
\end{equation}
denote the aggregate normalized
CSI feedback cost and \(\widetilde{B}_{\max}\) is the maximum
allowable normalized feedback budget. If
\(\widetilde{B}[t]>\widetilde{B}_{\max}\), the following penalty
is applied:
\begin{equation}
r_k[t] \leftarrow r_k[t] - \eta \frac{\widetilde{B}[t]-\widetilde{B}_{\max}}{K},
\end{equation}
where $\eta > 0$ controls the severity of the budget violation penalty.

\subsubsection{Optimization Objective}
The RL agent is trained to maximize the expected discounted sum of rewards
over a finite horizon:
\begin{equation}
\label{eq:return}
\max_{\pi} \;
\mathbb{E}_{\pi} \!\left[
\sum_{t=0}^{\infty} \gamma^{t}
\sum_{k=1}^{K} r_k[t]
\right],
\end{equation}
where $\pi$ denotes the control policy and $\gamma \in (0,1)$ is the discount factor. The discounted objective allows the agent to account for the long-term effects of CR selection on reconstruction quality, communication performance, feedback overhead, and switching frequency. The state variables enable the policy to respond to evolving channel and multi-user conditions, while the feedback and switching penalties discourage unnecessarily aggressive or frequent CR changes.

\section{Experimental Setup \& Training} \label{result}
\subsection{Dataset}
We consider a dataset generated through a time-domain simulation of a multi-user massive MIMO system compliant with 5G NR specifications at $3.5$~GHz using the MATLAB 5G Toolbox. The simulation models an outdoor cellular environment with $K = 2,4,6,8$ mobile users served by a base station equipped with a uniform planar array (UPA) of $N_t = 64$ antennas. Users move in a two-dimensional plane with time-varying velocities, and their positions are updated every slot of duration $0.5$~ms. The OFDM configuration employs $52$ resource blocks with a subcarrier spacing of $30$~kHz, yielding $\tilde{N}_s = 624$ subcarriers.

The channel impulse response (CIR) truncation length is determined from the average power delay profile of the generated channels. Since most of the channel energy is concentrated within the first $50$--$60$ delay taps, the CIR is truncated to $64$ taps to reduce dimensionality while preserving the dominant multipath components.

To capture diverse propagation conditions, three CDL channel families are instantiated for each user: CDL-A, CDL-C, and CDL-E, representing progressively more challenging propagation environments with different delay spreads and scattering characteristics. Perfect frequency-domain CSI is computed at every slot by combining time-varying path gains with the fixed path-delay filters extracted from the CDL channel objects.

The proposed dataset considers globally synchronized scenario transitions, where all users undergo the same high-level environmental change at the same time. This setting does not imply that users experience identical channel realizations; rather, it models a common propagation-regime shift across the network, while preserving user-specific differences in channel gains, mobility, fading realizations, SNR, and spatial correlation. Such common-mode transitions can occur in practical scenarios involving collective user mobility, e.g., passengers in a vehicle or train, network-wide environmental changes, large obstacles affecting a shared propagation region, or coordinated movement through different propagation zones. Under these conditions, the dominant propagation characteristics observed by multiple users may change in a correlated manner, leading to simultaneous transitions between regimes such as LoS, NLoS, and blockage.


The propagation process evolves through three distinct regimes:
\begin{itemize}
\item \textbf{Scenario 1 (easy regime):} highly correlated and low-rank channels favoring aggressive CSI compression,
\item \textbf{Scenario 2 (moderate regime):} intermediate scattering conditions with moderate spatial diversity,
\item \textbf{Scenario 3 (hard regime):} highly frequency-selective and distorted channels requiring lower compression.
\end{itemize}

The propagation process evolves through three distinct regimes corresponding to different channel conditions. Specifically, CDL-A represents the LoS (easy) regime, CDL-C represents the NLoS (moderate) regime, and CDL-E represents the blockage-dominated (hard) regime. Throughout the remainder of the paper, we refer to these regimes using the physical labels LoS, NLoS, and Blockage for clarity.


The generated dataset is stored as a five-dimensional tensor containing the frequency-domain CSI across subcarriers, antennas, users, and time slots, together with per-slot SNR values, user mobility information, scenario labels, and dynamic feedback-budget annotations. The complete dataset spans $8{,}000$ time slots and captures realistic temporal correlations in both small-scale and large-scale fading. For training and evaluation, $70\%$ of the dataset is used for training, while the remaining $30\%$ is reserved for testing. 

The source code for reproducing the experiments is available on GitHub \cite{ansarifard2026rlcsi}.
\subsection{CSI Normalization before feeding into AE}
The CSI feedback AE operates on real-valued inputs obtained by separating the complex delay-domain channel coefficients into their real and imaginary components, concatenated into a single vector of dimension $2 \times N_t \times N_s$. Normalization of this input vector is essential for stable and effective AE training. The delay-domain transformation via IFFT scales channel coefficients to magnitudes on the order of $10^{-7}$, far below the operating range of Tanh activation functions used throughout the encoder and decoder. 

A key challenge in this system is that the $K$ users experience statistically heterogeneous channel conditions, differing in path loss, multipath structure, and scenario exposure (LoS, NLoS, and blockage), making a single global normalization across all users inappropriate. Instead, we apply per-user scalar normalization using running mean and standard-deviation estimates. Unlike feature-wise normalization, this approach preserves the relative energy distribution across delay taps while accommodating heterogeneous user channel powers. The online update procedure is provided in Appendix~\ref{app:normalization}.
\subsection{Autoencoder design}
Each user is associated with a lightweight AE that maps the flattened real and imaginary components of the CSI matrix to a lower-dimensional latent representation and reconstructs it at the BS. The AE architecture is dynamically constructed based on the desired latent dimension $L$, which corresponds to a selected CR. Specifically, the encoder is composed of a sequence of fully connected layers that progressively reduce the feature dimension by a factor of two until reaching the target latent dimension. Each intermediate layer is followed by Layer Normalization and a ReLU activation to stabilize training and improve generalization. The decoder mirrors this structure in reverse, expanding the latent representation back to the original input dimension using fully connected layers with ReLU activations, except for the final output layer, which remains linear.

To ensure stable and reliable operation, a set of AEs is pretrained offline, each corresponding to a predefined CR. During pretraining, each AE is optimized independently using MSE loss over normalized CSI samples from the training dataset. This results in a bank of CR-specific models that provide consistent reconstruction performance and well-defined action outcomes for the RL agent.

During adaptive operation, the environment maintains $K$ independent AE banks (one per user), where each bank contains a small number of pretrained models corresponding to the available CR levels. At each time step, the RL agent selects a CR for each user, and the corresponding pretrained AE is activated for encoding and reconstruction. This design avoids the need for dynamically changing network architectures or transferring weights across different latent dimensions, thereby preserving stability in the learning process.

During online operation, all AE parameters remain frozen. The framework adapts to the observed channel and network conditions solely by switching the active pretrained encoder--decoder pair according to the CR selected by the RL policy.

\subsection{Double DQN agent}
We adopt a value-based deep RL approach using a Double DQN with a dueling network architecture. The dueling architecture decomposes the state--action value function as
\begin{equation}
Q(s,a;\theta)=V(s;\theta_v)+\left(A(s,a;\theta_a)
-\frac{1}{|\mathcal{A}|}\sum_{a'}A(s,a';\theta_a)\right),
\end{equation}
where $V(s;\theta_v)$ denotes the \emph{state-value function}, representing the overall quality of state $s$, and $A(s,a;\theta_a)$ denotes the \emph{advantage function}, which measures the relative importance of action $a$ compared to other actions in state $s$. The parameters $\theta_v$ and $\theta_a$ correspond to the value and advantage streams of the dueling network, respectively, and $\mathcal{A}$ denotes the action space. This decomposition improves representation learning and sample efficiency by disentangling state evaluation from action preference. 

Experience replay is employed to decorrelate training samples, and a target network is periodically updated to stabilize learning. During training, actions are selected using an $\epsilon$-greedy exploration policy with exponentially decaying $\epsilon$, while inference uses a fully greedy policy.

\subsection{Training Procedure}
The overall training procedure is summarized in Algorithm~\ref{alg:ae_rl}, where the AE is first pretrained over the training dataset, using normalized CSI to accelerate convergence. Subsequently, the RL agent interacts with the environment over the time series of CSI samples, selecting CRs for each user and receiving corresponding rewards. Switching to the pre-trained AE is triggered dynamically by the agent's actions. The RL agent is trained over multiple epochs using mini-batch gradient descent on replay-buffer transitions. Only the DQN parameters are updated during this stage; the pretrained AE parameters remain fixed. During deployment, the trained policy selects the CR and activates the corresponding encoder--decoder pair without online learning or model-parameter exchange. Although the proposed framework forms a closed-loop interaction between the UE and the BS, it does not incur continuous bidirectional learning traffic. The uplink transmission corresponds to standard CSI feedback using compressed representations, while the downlink signaling consists of low-rate control commands that update the compression configuration. The RL training and reward computation are performed entirely at the BS and do not require over-the-air learning parameter exchange.
\section{Performance Analysis}\label{analysis}
This section analyzes the performance of the proposed adaptive CSI compression framework based on joint RL and autoencoder design. We evaluate both the learning behavior during training and the generalization capability during evaluation on unseen CSI realizations.

\subsection{Training Performance and Policy Adaptation}

\subsubsection{Training Convergence and Scalability}

\begin{figure}[t]
    \centering
    \includegraphics[width=0.95\columnwidth]{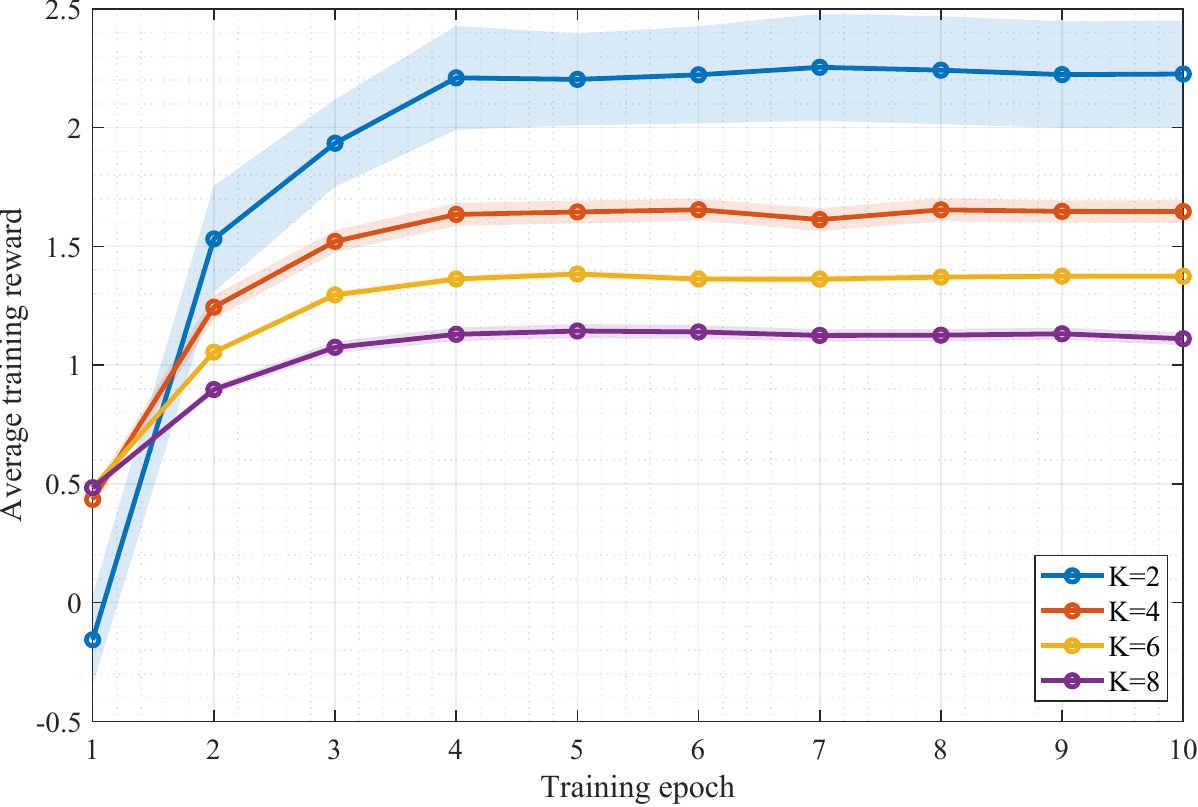}
    \vspace{-3mm}
    \caption{Average training reward across training epochs for different number of users.}
    \label{training_results}
\end{figure}

Fig.~\ref{training_results} illustrates the evolution of the average training reward for different numbers of users. Across all configurations, the reward increases rapidly during the initial training epochs, indicating that the RL agent quickly learns effective CR switching policies from the observed network state. After approximately four to five epochs, the reward converges and remains stable, suggesting that the policy has reached a near-stationary operating point.

An important observation is that the convergence behavior remains consistent across different network sizes. Although all configurations exhibit a similar learning trend, the achievable reward decreases as the number of users increases. This behavior is expected because larger multi-user systems impose stricter feedback-resource constraints and create a more challenging optimization problem. Specifically, the RL agent must balance CSI reconstruction quality, feedback overhead, switching cost, and communication performance across a larger set of users, resulting in a lower average reward despite successful policy convergence.

The fastest reward growth is observed during the first few training epochs, demonstrating that the proposed state representation provides sufficient information for efficient policy learning. The subsequent plateau indicates that additional training primarily refines the learned policy rather than producing substantial performance gains. Moreover, the relatively smooth convergence across all user configurations suggests that the proposed Dueling-DQN framework remains stable as the system dimension increases and does not require user-specific retraining.
\begin{algorithm}[t]
\caption{Adaptive CSI Compression with Double DQN and Pretrained AE Bank}
\label{alg:ae_rl}
\begin{algorithmic}[1]
\Require CSI dataset $\mathcal{H}$, pretrained AE bank $\{\mathrm{AE}^{(r)}\}_{r \in \mathcal{R}}$, 
dueling Q-network parameters $\theta$, target network parameters $\theta^{-}$,
learning rate $\alpha_{\mathrm{lr}}$, discount factor $\gamma$, CR set $\mathcal{R}$,
number of users $K$, number of time slots $T$

\State Initialize Q-network with parameters $\theta$
\State Initialize target network $\theta^{-} \leftarrow \theta$
\State Initialize replay buffer $\mathcal{D}$
\State Initialize normalization statistics $\mu_k[0], \sigma_k[0]$ for all users
\State Initialize active AE for each user: $\mathrm{AE}_k \leftarrow \mathrm{AE}^{(r_0)}$

\For{time slot $t = 1$ to $T$}
    \For{each user $k = 1$ to $K$}
        \State $H_{k,t} \gets$ CSI sample for user $k$ at time $t$
        \State $x_{k,t} \gets$ flattened real-imaginary CSI representation
        \Statex \hspace{2cm}of $H_{k,t}$

        \State Compute instantaneous statistics $\hat{\mu}_k[t], \hat{\sigma}_k[t]$
        \State Update normalization statistics:
        \Statex \hspace{1.5cm} $\mu_k[t] \gets (1 - \alpha^{\mathrm{m}}[t])\mu_k[t-1] + \alpha^{\mathrm{m}}[t]\hat{\mu}_k[t]$
        \Statex \hspace{1.5cm} $\sigma_k[t] \gets (1 - \alpha^{\mathrm{m}}[t])\sigma_k[t-1] + \alpha^{\mathrm{m}}[t]\hat{\sigma}_k[t]$

        \State Normalize input:
        \Statex \hspace{1.5cm} $\tilde{x}_{k,t} \gets \dfrac{x_{k,t} - \mu_k[t]}{\sigma_k[t]}$

        \State Observe state $s_{k,t}$ as in (\ref{state})
        \State Select action $a_{k,t} \in \mathcal{R}$ using an $\epsilon$-greedy policy

        \State Select the pretrained AE corresponding to 
        \Statex \hspace{1cm} the chosen CR: $\mathrm{AE}_k \leftarrow \mathrm{AE}^{(a_{k,t})}$
        
        \State Encode and reconstruct:
        \Statex \hspace{1.5cm} $\tilde{z}_{k,t}, \tilde{\hat{x}}_{k,t} \gets \mathrm{AE}_k(\tilde{x}_{k,t})$

        \State Denormalize reconstruction:
        \Statex \hspace{1.5cm} $\hat{x}_{k,t} \gets \sigma_k[t]\tilde{\hat{x}}_{k,t} + \mu_k[t]$

        \State Compute reconstruction error:
        \Statex \hspace{1.5cm} $e_{k,t} = \mathrm{NMSE}(\hat{x}_{k,t}, x_{k,t})$

        \State Compute reward $r_{k,t}$ as in (\ref{reward})
        \State Observe next state $s_{k,t+1}$
        \State Store transition $(s_{k,t}, a_{k,t}, r_{k,t}, s_{k,t+1})$ in $\mathcal{D}$
    \EndFor

    \If{$|\mathcal{D}| \geq B$}
        \State Sample mini-batch $\mathcal{B} \subset \mathcal{D}$
        \For{each transition $(s,a,r,s') \in \mathcal{B}$}
            \State $a^{*} \gets \arg\max_{a'} Q(s', a'; \theta)$
            \State $y \gets r + \gamma Q(s', a^{*}; \theta^{-})$
            \State $\mathcal{L}(\theta) \gets \left(y - Q(s,a;\theta)\right)^2$
        \EndFor
        \State Update Q-network:
        \Statex \hspace{1.5cm} $\theta \gets \theta - \alpha_{\mathrm{lr}} \nabla_{\theta} \mathcal{L}(\theta)$
    \EndIf

    \State Periodically update target network parameters $\theta^{-} \gets \theta$
\EndFor
\end{algorithmic}
\end{algorithm}

\subsubsection{Scenario-Aware Action Policy}

\begin{figure}[t]
	\centering
	\includegraphics[width=0.95\columnwidth]{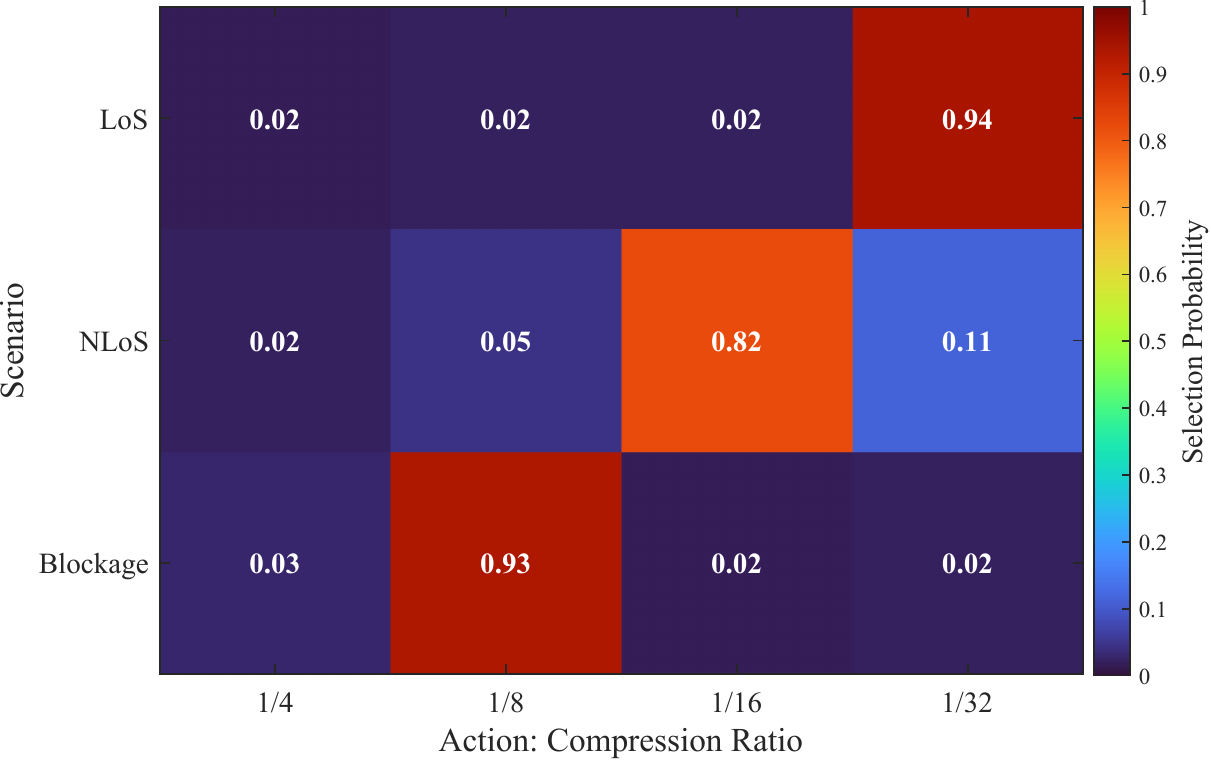}

   \vspace{-0.3em}
	\caption{Compression ratio selection probability per scenario for $K=4$ users. Each row corresponds to a channel scenario (LoS, NLoS, Blockage), and each column represents an action (compression levels).} 
	\label{action_h}
\end{figure}

Fig.~\ref{action_h} shows the CR selection probability of the trained RL policy under different propagation scenarios. The policy learns distinct switching behavior across scenarios. In the LoS case, the agent predominantly selects the strongest compression level, CR $=1/32$, with probability 0.94. This indicates that when the channel is more structured and easier to reconstruct, the policy reduces feedback overhead aggressively. For the NLoS scenario, the policy mainly selects CR $=1/16$, with probability 0.82, while occasionally selecting CR $=1/32$. This reflects a more conservative tradeoff between reconstruction accuracy and feedback reduction under less favorable channel conditions. In blockage conditions, the policy selects CR $ = 1/8$ with probability 0.93, showing that the agent increases the feedback dimension when channel reconstruction becomes more challenging. Therefore, the learned policy adapts the CR according to scenario difficulty: stronger compression is used in LoS, moderate compression in NLoS, and weaker compression in blockage.

\subsection{Baselines}
\subsubsection{Rule-Based Baseline}
To provide a non-learning benchmark, we design an \emph{Adaptive Threshold Heuristic (ATH)} that uses the same state information as the RL agent in \eqref{state}. As summarized in Algorithm~\ref{alg:ath}, ATH selects low-, medium-, or high-compression modes according to the instantaneous SINR, while using the smoothed NMSE and its temporal variation to account for reconstruction quality and channel drift. A safety override switches to a less aggressive compression mode when the NMSE exceeds $\theta_N$. Similarly, if the NMSE variation remains above $\theta_{\Delta}$ for $T_f$ consecutive slots, ATH adopts a more conservative mode to preserve CSI reconstruction fidelity.

\begin{algorithm}[t]
\caption{ATH Rule-Based CSI Compression}
\label{alg:ath}
\begin{algorithmic}[1]
    \For{each user $k$ at time $t$}
        \State Update $\overline{\mathrm{NMSE}}_k[t]$ and $\Delta \mathrm{NMSE}_k[t]$
        \If{$\mathrm{SINR}_k[t] > \theta_H$}
            \State $m_k[t] \gets \mathrm{HR}$
        \ElsIf{$\mathrm{SINR}_k[t] > \theta_M$}
            \State $m_k[t] \gets \mathrm{MR}$
        \Else
            \State $m_k[t] \gets \mathrm{LR}$
        \EndIf
        \If{\(\mathrm{NMSE}_k[t]>\theta_N\) and
        \(m_k[t]\neq\mathrm{LR}\)}
            \State \(m_k[t]\gets\mathrm{LR}\)
        \EndIf
        \If{$\Delta \mathrm{NMSE}_k[t] > \theta_{\Delta}$ for $T_f$ slots}
            \State $m_k[t] \leftarrow$ more conservative mode
        \EndIf
    \EndFor
\end{algorithmic}
\end{algorithm}

\subsubsection{Adaptive DNN-based CSI Feedback Baseline}
The adaptive DNN-based CSI feedback framework proposed in \cite{10012898} serves as the ideal benchmark to our proposed method. While it optimizes for local MSE, it remains "environmentally blind" to inter-user dynamics. By selecting this as a baseline, which we refer to as local-adaptive DNN (LAD) baseline, we can explicitly demonstrate the advantages of a system-level RL policy over a user-level greedy adaptive policy. By utilizing this baseline, we aim to illustrate that even a high-accuracy, DNN-based adaptive feedback system can suffer from performance degradation in multi-user interference scenarios if it lacks the global coordination and multi-objective optimization provided by our proposed framework.The LAD baseline is implemented as a learned local adaptive DNN-based CSI feedback scheme. 
For each candidate CR, a separate AE is trained, and an offline CR classifier is trained using labels obtained from the CR that provides the lowest reconstruction NMSE on the training data. 
During testing, the classifier predicts the CR for each user based on the current local CSI, and the corresponding pretrained AE is used for CSI reconstruction. 
Therefore, LAD approximates a local NMSE-driven CR selection rule, but it does not perform an exhaustive per-slot search over all CRs and does not explicitly optimize feedback overhead, SINR, or downlink sum rate.
\subsubsection{Fixed baselines}
For reference, several fixed CR baselines are also included in the comparison. These baselines use a single predefined CR throughout the entire transmission process and therefore do not perform any adaptive switching according to the channel condition. Specifically, fixed CR $=1/4$, CR $=1/8$, CR $=1/16$, and CR $=1/32$ configurations are evaluated. Each fixed baseline employs the corresponding pretrained autoencoder associated with its selected CR and keeps this configuration unchanged during all transmission slots. Consequently, these methods represent static CSI feedback strategies with different operating points on the reconstruction-versus-feedback-overhead tradeoff curve.

Lower compression levels (e.g., CR $=1/4$) preserve more CSI information and generally achieve better reconstruction quality, but at the expense of significantly larger feedback payloads. In contrast, higher compression levels (e.g., CR $=1/32$) greatly reduce the feedback overhead but suffer from increased reconstruction distortion under challenging propagation conditions. Since these baselines cannot adapt the CR according to the instantaneous channel scenario, their performance becomes highly sensitive to environmental changes. The fixed baselines therefore provide useful reference points for evaluating whether the proposed RL framework can learn an effective scenario-aware compression policy that dynamically balances CSI reconstruction quality and communication efficiency.
\begin{table}[t]
\centering
\caption{Main hyperparameters used for RL training and reward evaluation.}
\label{tab:hyperparameters}
\begin{tabular}{l c c}
\hline
Parameter & Symbol & Value \\
\hline
Rate weight & $\delta$ & $0.45$ \\
NMSE penalty weight & $\alpha$ & $2.4$ \\
Feedback-cost weight & $\beta$ & $0.20$ \\
Switching penalty & $c_{\mathrm{sw}}$ & $0.005$ \\
Budget-mismatch penalty & $\eta$ & $6.0$ \\
Bits per latent element & $q$ & $10$ \\
Candidate CRs & $\mathcal{R}$ & $\{1/4,1/8,1/16,1/32\}$ \\
Discount factor & $\gamma$ & $0.95$ \\
Initial exploration rate & $\epsilon_0$ & $0.8$ \\
Minimum exploration rate & $\epsilon_{\min}$ & $0.01$ \\
Exploration decay & -- & $0.9998$ \\
Learning rate & -- & $10^{-4}$ \\
Replay buffer size & -- & $2\times10^5$ \\
Mini-batch size & -- & $256$ \\
Target-network update interval & -- & $1000$ steps \\
\hline
\end{tabular}
\end{table}

The reward coefficients in Table~\ref{tab:hyperparameters} were selected using a validation set and then kept fixed for all reported experiments. The coefficients were chosen to balance the numerical scales of the physical metrics in the reward, so that the rate term, NMSE penalty, feedback-cost penalty, and switching penalty all contribute to the learned policy. No test-set information was used for selecting these values.
\subsection{Evaluation Performance}

All methods are evaluated under identical channel realizations, CSI samples, and physical-layer processing. The reconstructed CSI produced by each method is passed to the same ZF precoder, and the SINR, downlink sum rate, NMSE, normalized CSI feedback cost, and reward are computed using the same formulations. Therefore, the observed differences are caused by the CR selection policy and the corresponding pretrained AE used for CSI reconstruction. The discussion below emphasizes physical communication and feedback metrics, namely downlink sum rate, NMSE, CSI feedback cost, and recovery behavior after scenario transitions. The reward is reported as a secondary aggregate metric, since it combines these quantities according to the RL training objective.

\subsubsection{Downlink Sum Rate versus CSI Feedback}

\begin{figure}[t]
	\centering
	\includegraphics[width=0.90\columnwidth]{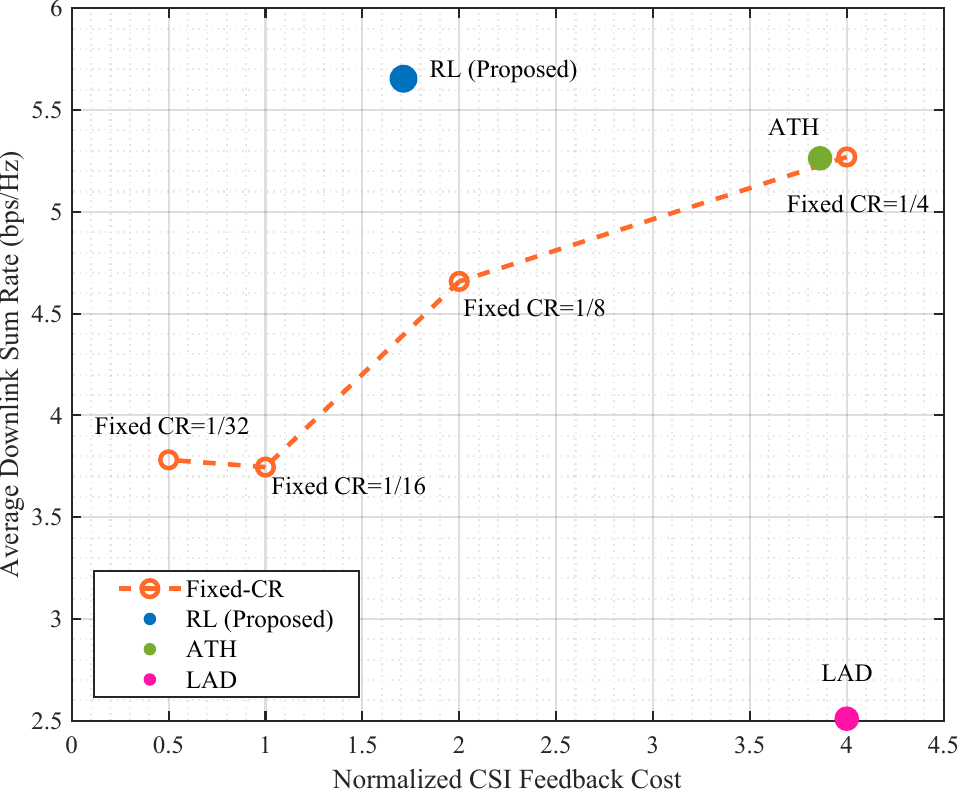}
	\caption{Downlink sum rate versus normalized CSI feedback cost
for \(K=4\) users.} 
	\label{rate_dist}
\end{figure}

Fig.~\ref{rate_dist} illustrates the tradeoff between the downlink sum rate and the normalized CSI feedback cost for $K=4$ UEs. The latter represents the compressed CSI payload transmitted from the UEs to the BS relative to the reference configuration with CR $=1/4$. The aggregate normalized feedback cost at slot $t$ is defined in \eqref{total cost}, and its average over $T$ evaluation slots is $
\overline{\widetilde{B}}
=
\frac{1}{T}
\sum_{t=1}^{T}
\widetilde{B}[t].
$

The fixed-CR baselines exhibit the expected static rate--feedback tradeoff. A larger latent representation, such as CR $=1/4$, preserves more CSI information and generally improves the reconstructed channel used for ZF precoding, but incurs greater feedback overhead. Conversely, more aggressive compression reduces the feedback payload at the cost of higher reconstruction distortion and potentially lower downlink sum rate. Each fixed-CR baseline therefore represents a single operating point on the rate--feedback curve.

The proposed RL framework achieves a more favorable operating point by dynamically selecting among pretrained compression-specific AEs according to the observed channel and network conditions. It can employ a larger latent representation when CSI accuracy is important and switch to more aggressive compression when reducing feedback is more beneficial. Relative to the fixed CR $=1/4$ baseline, the proposed method reduces the average normalized feedback cost from $4$ to $1.7$, corresponding to a $57.5\%$ reduction, while improving the average downlink sum rate by $5.66\%$. Compared with ATH, it reduces the feedback cost by $55.84\%$ and improves the sum rate by $5.66\%$. Compared with LAD, it reduces the feedback cost by $57.5\%$ and improves the sum rate by $124\%$. These results demonstrate that the proposed framework improves the physical rate--feedback tradeoff rather than merely maximizing the scalar reward.

LAD incurs relatively high feedback overhead without achieving a corresponding sum-rate gain because its compression decisions are driven primarily by reconstruction-oriented labels. Consequently, it does not explicitly account for the effects of CSI errors on multi-user ZF precoding, residual interference, and the resulting communication rate. A larger latent representation or lower local NMSE therefore does not necessarily provide better system-level performance. In contrast, the proposed communication-aware RL formulation learns the joint impact of CSI accuracy, feedback overhead, and downlink rate.

\begin{figure*}[!ht]
    \centering

    \subfloat[]{%
        \includegraphics[width=0.32\textwidth]{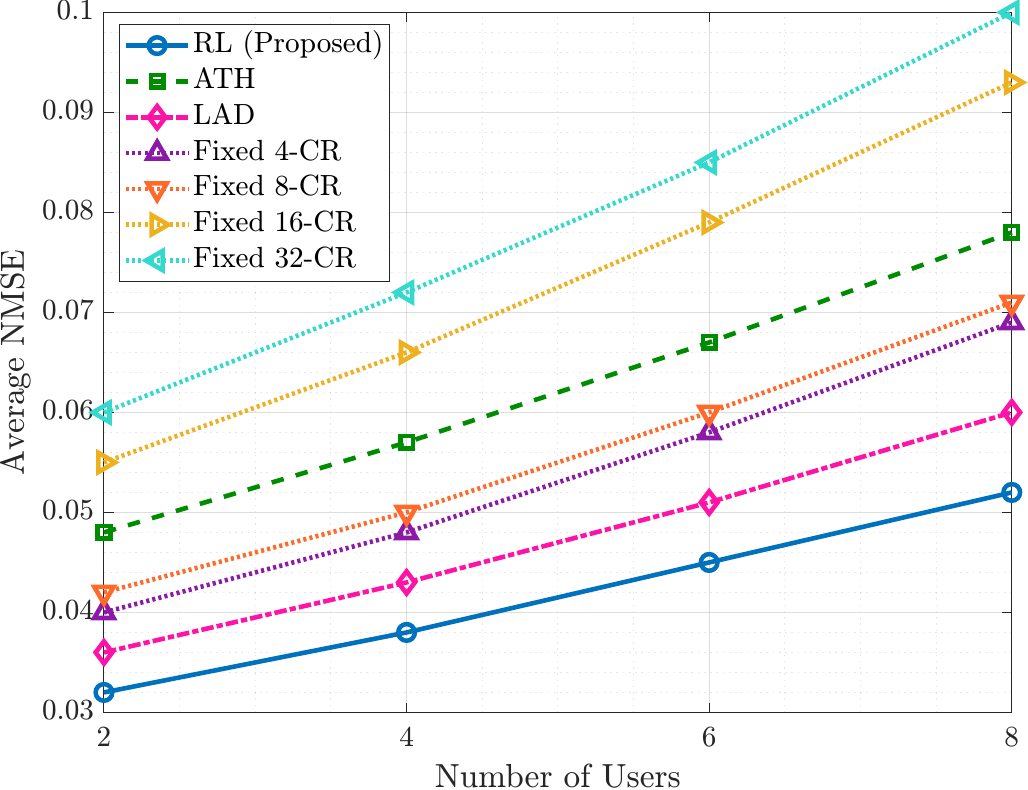}
        \label{users_nmse}
    }
    \hfill
    \subfloat[]{%
        \includegraphics[width=0.32\textwidth]{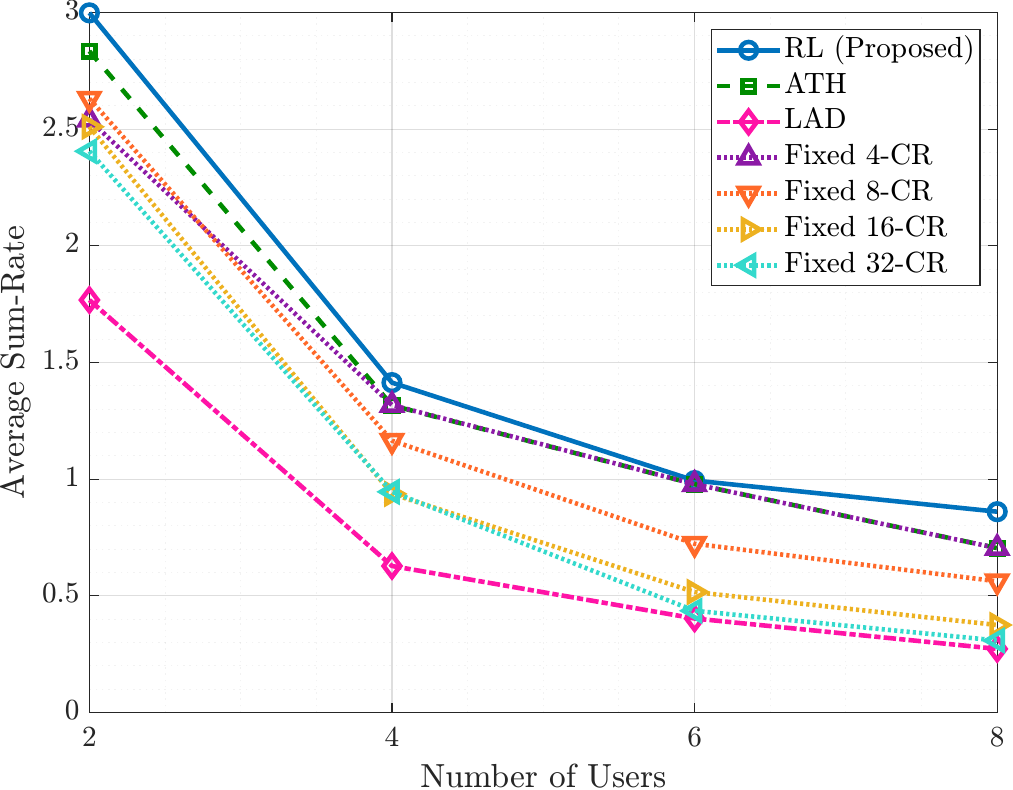}
        \label{users_rate}
    }
    \hfill
    \subfloat[]{%
        \includegraphics[width=0.32\textwidth]{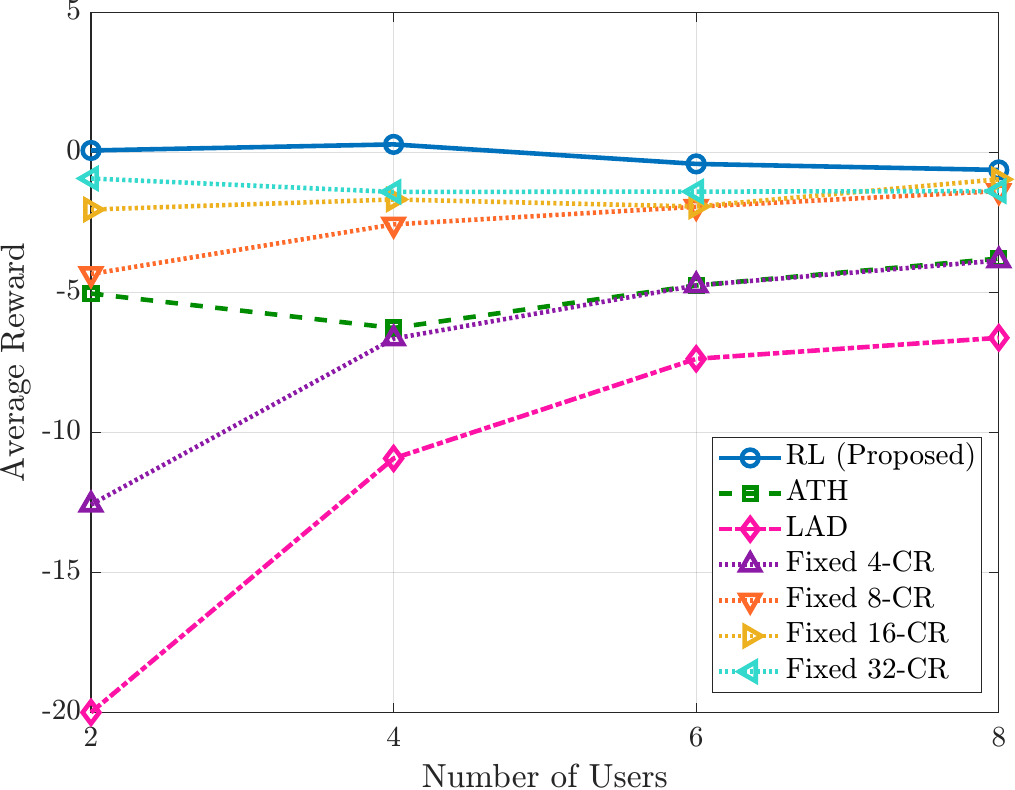}
        \label{users_reward}
    }
    \vspace{-0.3em}
    \caption{Performance comparison versus the number of users. (a) Average NMSE, (b) average downlink sum rate, and (c) average scalar reward utility. The reward summarizes the weighted rate--accuracy--feedback tradeoff and is not interpreted alone as a physical-layer performance metric.} 
    \label{sca}
\end{figure*}

\subsubsection{Scalability with the Number of Users}

Fig.~\ref{sca} evaluates scalability in terms of average NMSE, downlink sum rate, and reward as the number of users increases. As shown in Fig.~\ref{users_nmse}, the proposed RL framework consistently achieves the lowest NMSE by adapting the compression level to the observed system conditions. Fig.~\ref{users_rate} further shows that it maintains the highest downlink sum rate, indicating that the reconstructed CSI preserves the information required for effective ZF precoding and interference suppression.

Fig.~\ref{users_reward} reports the corresponding average reward, which combines rate, reconstruction error, feedback cost, and switching cost; therefore, a method can obtain a relatively high reward by strongly reducing CSI feedback overhead even if its NMSE or sum rate is worse. This explains why some aggressive fixed-CR baselines, especially CR=$1/16$ and CR=$1/32$, approach the proposed RL framework in average reward as the number of users increases. These fixed schemes use small latent representations and do not incur switching penalties, which improves their scalar reward under the selected utility weights. However, this should not be interpreted as comparable communication performance, since the same fixed schemes exhibit worse NMSE and lower downlink sum rate. Overall, the proposed RL framework provides a more balanced operating point, achieving lower reconstruction distortion and higher downlink sum rate while the reward trend serves only as an aggregate confirmation of the rate--accuracy--overhead tradeoff.

The main performance gains of the proposed framework over the considered baselines are summarized in Table~\ref{tab:headline_gains}.
\subsubsection{Scenario Transitions}

\begin{figure*}[!ht]
    \centering

    \subfloat[]{%
        \includegraphics[width=0.32\textwidth]{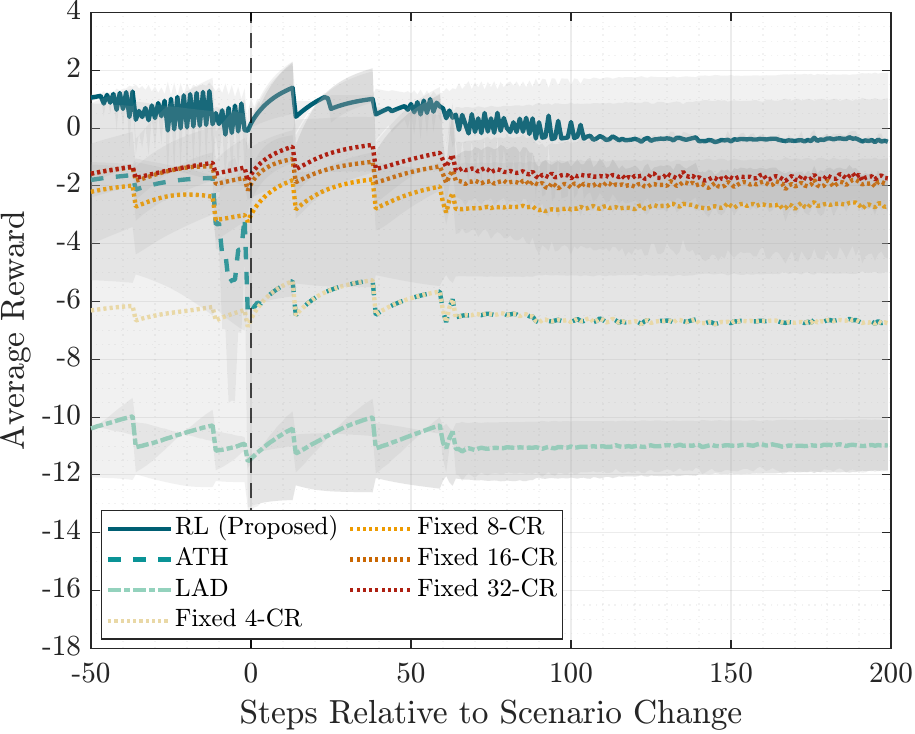}
        \label{fig:eval_a}
    }
    \hfill
    \subfloat[]{%
        \includegraphics[width=0.32\textwidth]{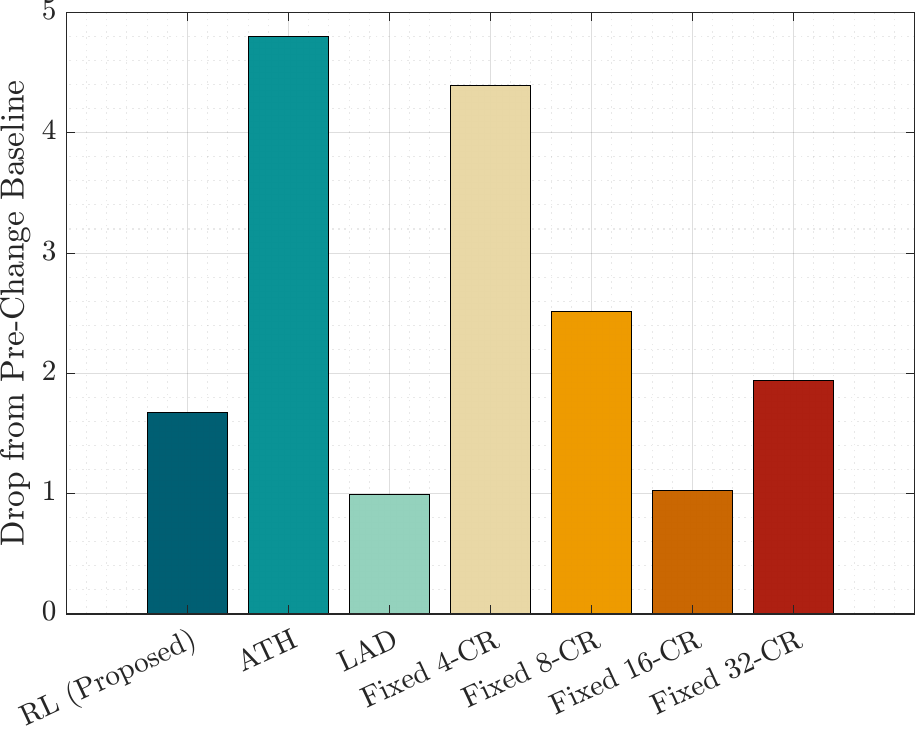}
        \label{fig:eval_b}
    }
    \hfill
    \subfloat[]{%
        \includegraphics[width=0.32\textwidth]{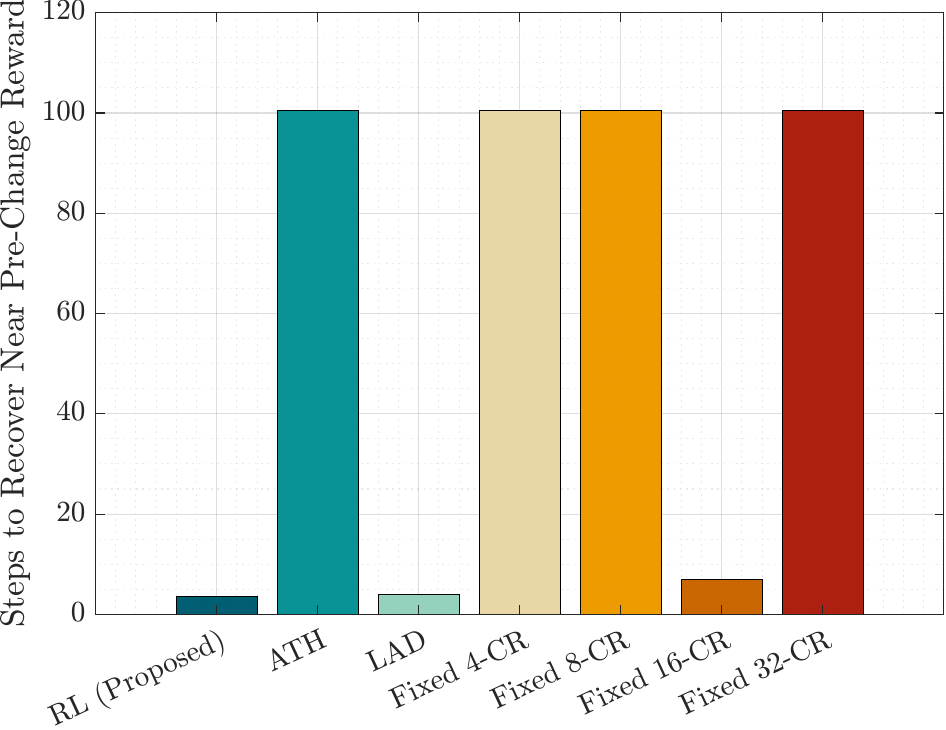}
        \label{fig:eval_c}
    }

    \vspace{-0.3em}
    \caption{Scenario-transition adaptation performance of the proposed RL framework and baseline methods for $K=4$ users using a 200-slot transition window. 
    (a) Evolution of the aggregate transition metric around the scenario change, where slot zero denotes the transition instant. 
    (b) Immediate post-transition degradation relative to the pre-transition operating level. 
    (c) Recovery time measured as the number of slots required to return close to the pre-transition operating level.}
    
    \label{fig:evaluation}
\end{figure*}

Fig.~\ref{fig:evaluation} evaluates the adaptation capability of the compared methods around channel-scenario transitions. The horizontal axis denotes the transmission slots relative to the transition instant, with zero marking the propagation change. All methods use the same trained policies and evaluation trajectory; the displayed window only selects the plotted portion and does not affect their decisions. The main metrics are the immediate post-transition degradation and the recovery time, which reflect sensitivity to the channel change and the speed of restoring the communication--feedback operating point. The transition metric summarizes the combined effects of downlink rate, NMSE, feedback cost, and switching cost.

Fig.~\ref{fig:eval_a} shows the behavior over the 200-slot window. The proposed RL framework experiences smaller degradation and stabilizes more rapidly than the baselines. This indicates that the learned policy can switch among the pretrained AEs after a propagation change, limiting reconstruction degradation and preserving communication performance without online retraining.

ATH relies on manually selected thresholds, which may become temporarily mismatched to the new channel statistics and cause larger transient degradation. LAD can respond more smoothly through its learned classifier, but its decisions remain local and reconstruction-oriented. Since it does not explicitly optimize the long-term communication--feedback tradeoff, it does not achieve the same post-transition operating level as RL.

The fixed-CR baselines cannot adapt their CSI feedback dimension after a transition. A less compressed configuration preserves CSI quality at a constant feedback cost, whereas a more aggressive configuration reduces overhead but may incur greater reconstruction error. Any eventual stabilization of these baselines reflects the channel reaching a new steady regime rather than active adaptation.

Fig.~\ref{fig:eval_b} quantifies the immediate degradation relative to the pre-transition operating level. The proposed RL framework achieves one of the smallest drops, demonstrating greater robustness than static compression and hand-crafted adaptation. Fig.~\ref{fig:eval_c} reports the recovery time, showing that RL returns close to its pre-transition operating level within fewer slots than ATH and several fixed-CR baselines. Thus, the proposed method improves not only the average operating point but also the resilience to abrupt channel changes.

\begin{table}[t]
\centering
\caption{Average quantitative gains of the proposed RL framework over $K=\{2,4,6,8\}$ users.}
\label{tab:headline_gains}
\begin{tabular}{l c}
\hline
Metric & Average gain of RL \\
\hline
Feedback-cost reduction over ATH & $51.62\%$ \\
Feedback-cost reduction over LAD & $53.08\%$ \\
Feedback-cost reduction over Fixed CR=$1/4$ & $55.50\%$ \\
\hline
Average sum-rate gain over ATH & $9.24\%$ \\
Average sum-rate gain over LAD & $139.40\%$ \\
Average sum-rate gain over best fixed-CR baseline & $12.28\%$ \\
\hline
NMSE reduction over ATH & $33.21\%$ \\
NMSE reduction over LAD & $11.96\%$ \\
NMSE reduction over best fixed-CR baseline & $21.97\%$ \\
\hline
\end{tabular}
\end{table}
\begin{table*}[t]
\centering
\caption{Comparison of CSI feedback methods in terms of online modules, model size, and inference latency. Decision latency refers to compression selection overhead, while total latency includes full CSI reconstruction for all users.}
\label{tab:method_comparison}
\small 

\begin{tabular}{l l c c c}
\hline
\textbf{Method} & \textbf{Online module} & \textbf{Params} & \textbf{Latency} & \textbf{Real Time} \\
\hline
Fixed CR & AE enc--dec & $\sim$1--2M &116.91 ms (total) & \checkmark \\

ATH baseline & Rule-based selection & -- & 1.8 $\mu$s (decision) & \checkmark \\

\multirow{2}{*}{LAD baseline~\cite{10012898}} 
& Multiple AEs & $\sim$4--8M & 917.79 $\mu$s (total) & \checkmark \\
& + threshold logic & & (incl. selection) & \\

\textbf{RL (Ours)} 
& \textbf{MLP policy (Inference mode)} 
& \textbf{$\sim$40k} 
& \textbf{50.12 $\mu$s (decision)} 
& \textbf{\checkmark} \\

RL (full pipeline) 
& Policy + AE enc--dec 
& $\sim$40k + 1--2M 
& 250.08 ms (total) 
& \checkmark \\

RL training network 
& Dueling DQN (Policy + Target Network) 
& $\sim$200k 
& 245 $\mu$s 
& Offline \\

\hline
\end{tabular}
\end{table*}

\subsection{Complexity}
While RL is often associated with high computational complexity, the proposed framework decouples the resource-intensive learning phase from real-time operation. As summarized in Table~\ref{tab:method_comparison}, we categorize the RL system's footprint into three distinct operational states: the RL training network, the online RL policy, and the RL full pipeline.

The RL training network utilizes a Dueling DQN architecture ($\approx 200$k parameters) to learn the optimal control logic. However, this complexity is strictly confined to the offline phase. During online deployment, the agent transitions into a lightweight inference mode. By executing only a single forward pass of the multilayer perceptron (MLP)-based feature extractor and advantage head, the agent selects a compression action in just $\approx 50.12~\mu$s per user. This decision latency represents a negligible overhead, less than $5\%$ of the typical $1$~ms CSI update interval, ensuring that the added intelligence does not bottleneck the physical layer. The RL full pipeline entry accounts for the total system latency  measured in our simulation environment. While the raw reconstruction time in a software-based implementation reaches $\approx 250 ~\mathrm{ms}$, this process is highly parallelizable in dedicated BS hardware. The critical metric for real-time control is the RL decision latency ($50.12~\mathrm{\mu s}$), which ensures that the intelligence required to adapt the CSI CR occurs well within the $1$~ms transmission time interval (TTI) constraint for real-time FDD MIMO systems.

In contrast to the ATH baseline, which relies on rigid rule-based selection, or the LAD baseline, which requires maintaining multiple specialized AEs ($\approx 4$--$8$M parameters), our RL approach provides a scalable and memory-efficient solution. It achieves a superior balance: it is nearly as fast as simple rule-based logic while providing the environment-adaptive intelligence required to navigate the trade-off between feedback bit-budgets and reconstruction accuracy as channel conditions drift. Although maintaining multiple AEs per user introduces additional memory requirements, the overall storage overhead remains modest. The number of CR levels is small (e.g., $3–5$), and each AE has a compact fully connected architecture with a limited number of parameters. Consequently, the total memory footprint scales linearly with the number of CRs and remains on the order of a few megabytes, which is negligible for modern base station hardware. Moreover, all models are stored at the BS, where memory and computational resources are significantly less constrained than at the user equipment. This design therefore provides a practical and scalable solution that balances memory usage, stability, and adaptive performance.

\section{Conclusion}\label{conclusion}

This paper proposed a system-aware adaptive CSI compression framework for multi-user MIMO systems based on RL-guided switching among pretrained AEs. Unlike adaptive CSI feedback approaches that require online AE retraining or weight updates, the proposed framework dynamically selects among compression-specific pretrained AEs according to the observed channel and network conditions. This enables rapid CR adaptation while keeping all AE parameters frozen, thereby reducing online computational complexity and avoiding encoder--decoder synchronization problems.

The proposed architecture jointly considers CSI reconstruction quality, feedback overhead, switching cost, and achievable downlink sum rate through sequential decision-making. The RL agent learns a scenario-aware compression strategy capable of adapting to changing propagation conditions, including LoS, NLoS, and blockage scenarios. Experimental results demonstrated that the proposed framework improves the physical communication--feedback tradeoff compared with rule-based, supervised, and fixed CR baselines.


The results further showed that the proposed RL-guided switching framework achieves faster recovery after scenario transitions while maintaining stable long-term performance under non-stationary wireless environments. In addition, the learned policy provides a favorable rate--feedback tradeoff by dynamically balancing CSI fidelity and feedback efficiency according to the instantaneous channel condition. The advantage of the proposed method remains consistent as the number of users grows, highlighting its scalability for dense multi-user MIMO systems.


\appendices
\section{Online CSI Normalization Details} \label{app:normalization}
To handle the temporal non-stationarity introduced by smooth channel drift across scenarios, we further adopt an online adaptive normalization strategy. Static normalization statistics computed on training data become mismatched when the channel conditions evolve at test time. We therefore maintain a running estimate of $\mu_k$ and $\sigma_k$ for each user using  exponential moving average (EMA) updates:
\begin{equation}
\left\{
\begin{aligned}
    \mu_k[t] &= \left(1 - \alpha^{\text{m}}[t]\right)\mu_k[t-1] + \alpha^{\text{m}}[t]\hat{\mu}_k[t], \\
    \sigma_k[t] &= \left(1 - \alpha^{\text{m}}[t]\right)\sigma_k[t-1] + \alpha^{\text{m}}[t]\hat{\sigma}_k[t],
\end{aligned}
\right.
\end{equation}
where $\hat{\sigma}_k[t]$ is the instantaneous standard deviation of the current CSI sample and $\alpha^{\text{m}}[t]$ is an adaptive momentum coefficient. To ensure rapid adaptation at the beginning of each episode, where the mismatch between initialized statistics and current channel conditions is largest, we use a decaying momentum schedule:
\begin{equation}
    \alpha^{\text{m}}[t] = \max\!\left(\alpha_{\min}^{\text{m}},\ \frac{1}{t+1}\right).
\end{equation}

This sets $\alpha^{\text{m}}[0] = 1.0$, so the normalizer fully adopts the statistics of the first observed sample, and decays toward a stable floor $\alpha_{\min}^{\text{m}} = 0.01$ after approximately 100 time slots. This design ensures that the normalizer converges quickly at episode onset while remaining stable and drift-tracking during long rollouts. The same online statistics are used consistently for both the forward normalization passed to the autoencoder and the inverse denormalization applied to the reconstructed output before SINR computation, ensuring no scale mismatch is introduced between the compressed feedback and the physical channel estimate used at the base station.

\bibliographystyle{IEEEtran}
\bibliography{Bibliography}
\ifCLASSOPTIONcaptionsoff
  \newpage
\fi



\vfill


\end{document}